\documentclass[preprint,aps,pre,longbibliography]{revtex4-1}
\usepackage[utf8]{inputenc}
\usepackage{epsfig}
\usepackage{subfigure}
\usepackage{latexsym}
\usepackage{amsmath}
\usepackage{psfrag}
\usepackage{fullpage}
\usepackage{graphicx}
\usepackage{amssymb}
\usepackage{color}
\usepackage{placeins}

\newcommand{\avg}[1]{\left\langle #1\right\rangle}

\newcommand{\T}[1]{\textrm{#1}}

\newcommand{\M}[1]{{\bf #1}}
\newcommand{\V}[1]{{\bf #1}}

\begin{document}
 \title{Controllability of {multiplex}, multi-timescale networks}

\author{M\'arton P\'osfai}
\email{posfai@ucdavis.edu}
\affiliation{Complexity Science Center and Department of Computer Science, University of California, Davis, CA 95616, USA}
\affiliation{Department of Physics of Complex Systems, E\"otv\"os University, Budapest, H-1117, Hungary}
\author{Jianxi Gao}
\affiliation{Center for Complex Network Research, Department of Physics, Northeastern University, Boston, MA 02115, USA}
\author{Sean P. Cornelius}
\affiliation{Center for Complex Network Research, Department of Physics, Northeastern University, Boston, MA 02115, USA}
\author{Albert-L\'aszl\'o Barab\'asi}
\affiliation{Center for Complex Network Research, Department of Physics, Northeastern University, Boston, MA 02115, USA}
\affiliation{Center for Cancer Systems Biology, Dana-Farber Cancer Institute, Harvard University, Boston, MA 02215, USA}
\affiliation{Department of Medicine, Brigham and Women’s Hospital,
Harvard Medical School, Boston, MA 02115, USA}
\affiliation{Center for Network Science, Central European University, Budapest, H-1051, Hungary}
\author{Raissa M. D'Souza}
\affiliation{Complexity Science Center, Department of Computer Science and Department of Mechanical and Aerospace Engineering, University of California, Davis, CA 95616, USA}
\affiliation{{Santa Fe Institute, 1399 Hyde Park Road, Santa Fe, NM 87501,  USA}}

\date{\today}

\begin{abstract}
The paradigm of layered networks is used to describe many real-world systems -- from biological networks, to social organizations and transportation systems.  While recently there has been much progress in understanding the general properties of multilayer networks, our understanding of how to control such systems remains limited.  One fundamental aspect that makes this endeavor challenging is that each layer can operate at a different timescale, thus we cannot directly apply standard ideas from structural control theory of individual networks.  Here we address the problem of controlling multilayer and multi-timescale networks {focusing on two-layer multiplex networks with one-to-one interlayer coupling. We investigate} the practically relevant case when the control signal is applied to the nodes of one layer. We develop a theory based on disjoint path covers to determine the minimum number of inputs ($N_\T i$) necessary for full control. 
We show that if both layers operate on the same timescale then the network structure of both layers equally affect controllability.
 In the presence of timescale separation, controllability is enhanced if the controller interacts with the faster layer: $N_\T i$ decreases as the timescale difference increases up to a critical timescale difference, above which $N_\T i$ remains constant and is completely determined {by} the faster layer. We show that the critical timescale difference is large if Layer I is easy and Layer II is hard to control in isolation.
  In contrast, control becomes increasingly difficult if the controller interacts with the layer operating on the slower timescale and increasing timescale separation leads to increased $N_\T i$, again up to a critical value, above which $N_\T i$ still depends on the structure of both layers. This critical value is largely determined by the longest path in the faster layer that does not involve cycles.
{By identifying the underlying mechanisms that connect timescale difference and controllability for a simplified model, we provide crucial insight into disentangling how our ability to control real interacting complex systems is affected by a variety of sources of complexity.}
\end{abstract}

\maketitle


\section{Introduction}\label{sec:introduction}

Over the past two decades, the theory of networks proved to be a powerful tool for understanding individual complex systems~\cite{ALB02,NEW03}.  However, it is now increasingly appreciated that complex systems do not exist in isolation, but interact with each other~\cite{KIV14, BOC14}. Indeed, an array of phenomena -- from cascading failures~\cite{BUL10,BRU12} to diffusion~\cite{GOM13} -- can only be fully understood if these interactions are taken into account. Traditional network theory is not sufficient to describe the structure of such systems, so in response to this challenge, the paradigm of multilayer networks is being actively developed. Here we study a fundamental, yet overlooked aspect of multilayer networks: each individual layer can operate at a different timescale. Particularly, we address the problem of controlling multilayer, multi-timescale systems {focusing on two-layer multiplex networks}. Recently significant efforts have been made to uncover how the underlying network structure of a system affects our ability to influence its behavior~\cite{WAN02,SOR07,LIU11,WAN12,YUA13,COR13,POS13,GAO14,IUD15}. However, despite the appearance of coupled systems from infrastructure to biology, the existing literature -- with a few notable exceptions~\cite{CHA14,MEN16,YUA14,ZHA16} -- has focused on control of networks in isolation, and the role of timescales remains unexplored.

Control of multilayer networks is important for many applications. For example, consider a CEO aiming to lead a company consisting of employees and management. Studying the network of managers or the network of employees in isolation does not take into account important interactions between the different levels of hierarchy of the company. On the other hand, treating the system as one large network ignores important differences between the dynamics of the different levels, e.g. management may meet weekly, while employees are in daily interaction. In general, the interaction of timescales plays an important role in organization theory~\cite{ZAH99}. Or consider gene regulation in a living cell. External stimuli activate signaling pathways which through a web of protein-protein interactions affect transcription factors responsible for gene expression.  The activation of a signaling pathway happens on the timescale of seconds, while gene expression typically takes hours~\cite{ALB02b}. As a third example, consider an operator of an online social network who wants to enhance the spread of certain information by interacting with its users. However, a user may subscribe to multiple social networking services and may opt to share news encountered in one network through a different one -- out of reach of the operator. The dynamics of user interaction on different websites can be very different depending on user habits and the services offered~\cite{LER10,KWA10,BAK12}. For example the URL shortening service Bit.ly reports that the half-life of shared links depends on the social networking platform used: half the clicks on a link happened within 2.8 hours after posting on Twitter, within 3.2 hours on Facebook and within 7.4 hours on Youtube~\cite{BIT11}.

Common features of these examples are that (i)~each interacting subsystem is described by a separate complex network; (ii)~the dynamics of each subsystem operate on a different, but often comparable timescale and (iii)~the external controller directly interacts with only one of the subsystems. Here we study the control properties of a model that incorporates these common features, yet remains tractable. {More specifically, we study discrete-time linear dynamics on two-layer multiplex networks, meaning that we assume one-to-one coupling between the nodes of the two layers. This choice ensures both analytical tractability and the isolation of the role of timescales from the effect of more complex multilayer network structure. Identifying the underlying mechanisms that govern the controllability of this simple model provides crucial insight into disentangling how our ability to control real interacting complex systems is affected by a variety of sources of complexity.}

So far only limited work investigated {controllability} of multilayer networks. Menichetti et al. investigated the controllability of two-layer multiplex networks governed by linear dynamics such that the dynamics of the two layers are not coupled, but the input signals in the two layers are applied to the same set of nodes~\cite{MEN16}. Yuan et al. identified the minimum number of inputs necessary for full control of diffusion dynamics, allowing the controller to interact with any layer~\cite{YUA14}. Zhang et al. investigated the controllable subspace of multilayer networks with linear dynamics without timescale separation if the controller is limited to interact with only one layer; showing that it is more efficient to directly control peripheral nodes than central ones~\cite{ZHA16}. Here we also limit the controller to one layer, yet by exploring the minimum input problem, we offer a direct metric which allows us to compare our findings to previous results for single-layer networks~\cite{LIU11}. More so, the key innovation of our work is that we take into account the timescale of the dynamics of each layer, a mostly overlooked aspect of multilayer networks.

It is worth mentioning the recent work investigating the related, but distinct problem of controllability of networks with time-delayed linear dynamics~\cite{QIA16}. The key difference between time-delay and timescale difference is that for time-delayed dynamics the state of a node will depend on some previous state of its neighbors; however, the typical time to change the state of a node remains the same throughout the system. While in case of timescale difference, the typical time needed for changes to happen can be different in different parts of the system.

In the next section, we introduce a simple model that captures some common properties of multilayer networks and we describe the problem setup. In Sec.~\ref{sec:min_input_prob}, we develop a theory to determine the minimum number of inputs required for controlling multiplex, multi-timescale networks with discrete-time linear dynamics relying on graph combinatorial methods. In Sec.~\ref{sec:results}, we use networks with tunable degree distribution to systematically uncover the role of network structure and timescale separation. We study three scenarions: no timescale separation, Layer I operates faster and Layer II operates faster. Finally, in Sec.~\ref{sec:conclusions} we provide a discussion of our results and we outline open questions.

\section{Model definition}\label{sec:model_definition}

We aim to study the controllability of coupled complex dynamical systems with the following properties: (i) each subsystem (layer) is described by a complex network; (ii) the operation of each layer is characterized by a different timescale and (iii) the controller only interacts directly with one of the layers. We propose a model that satisfies these requirements and yet is simple enough to remain tractable. We focus on two-layer {multiplex} systems, meaning that there is a one-to-one correspondence between the nodes of the two layers.

The model is defined by a weighted directed two-layer multiplex network $\mathcal M$ which consists of two networks $\mathcal L_\T{I}$ and $\mathcal L_\T{II}$ called layers and a set of links $E_{\T{I},\T{II}}$ connecting the nodes of the different layers. Each layer $\mathcal L_\alpha$ (where $\alpha\in\{\T{I},\T{II}\}$) consists of a set of nodes $V_\alpha=\{v_1^\alpha,v_2^\alpha,\ldots,v_N^\alpha\}$ and a set of links $E_\alpha$, where a directed link $(v_i^\alpha,v_j^\alpha,w_{ij}^\alpha)\in E_\alpha$ is an ordered node pair and a weight representing that node $v_i^\alpha$ influences node $v_j^\alpha$ with strength $w^\alpha_{ij}$. The two layers are connected by link set $E_{\T{I},\T{II}}=\{(v_i^\T{I},v_i^\T{II}, w_i^\T{I,II})\vert i=1,2\ldots,N\}$, in other words, there is directed one-to-one coupling from Layer I to Layer II (Fig.~\ref{fig:3examples}a). Although the links are weighted, the exact values of the weights do not have to be known for our purposes.

Our goal is to control the system by only interacting directly with Layer I. We study linear discrete-time dynamics
\begin{equation}\label{eq:dynamics}
\begin{aligned}
\V{x}_\T{I}(t) &= \M{A}_\T{I}\V{x}_\T{I}(t-\tau_\T{I}) + \M{B}\V{u}(t-\tau_\T{I})\\
\V{x}_\T{II}(t) &= \M{A}_\T{II}\V{x}_\T{II}(t-\tau_\T{II}) + \Delta_{\tau_\T{I}}(t)\M{D}\V{x}_\T{I}(t - \tau_\T{I})
\end{aligned}
\begin{aligned}
\quad\text{ if }&(t\bmod\tau_\T{I})=0,\\
\quad\text{ if }&(t\bmod\tau_\T{II})=0,
\end{aligned}
\end{equation}
where $\V x_\T{I}(t)$ and $\V x_\T{II}(t)\in \mathbb{R}^N$ represent the state of nodes in Layer I and II; the matrices $\M{A}_\T{I}$ and $\M{A}_\T{II}\in \mathbb{R}^{N\times N}$ are the transposed weighted adjacency matrices of Layer I and II, capturing their internal dynamics. The weighted diagonal matrix $\M D\in \mathbb{R}^{N\times N}$ captures how Layer I affects Layer II. 

Vector $\V u(t)\in \mathbb R^M$ provides the set of independent inputs and the matrix $\M B\in \mathbb{R}^{N\times M}$ defines how the inputs are coupled to the system. To differentiate between the function $\V u(t)$ and an instance of the function at a given time step, we refer to a component $u_i(t)$ of vector $\V u(t)$ as an independent input, and we call its value at time step $t^\prime$, $u_i(t=t^\prime)$, a signal.

Finally, $\tau_\T{I}, \tau_\T{II}\in \{1,2,\ldots\}$ are the timescale parameters of each subsystem, meaning that the state of Layer I is updated according to Eq.~(\ref{eq:dynamics}) every $\tau_\T{I}$th time step; and Layer II is updated every $\tau_\T{II}$th time step. And
\begin{equation}
\Delta_{\tau_\T{I}}(k) = \left\{ \begin{aligned}
1 \quad\text{ if }&(k\bmod \tau_\T{I})=0,\\
0 \quad\text{ if }&(k\bmod \tau_\T{I})\neq 0,
\end{aligned}\right.
\end{equation}
is the Kronecker comb, meaning that Layer I directly impacts the dynamics of Layer II if the two layers simultaneously update. We investigate three scenarios: (i) the subsystems operate on the same timescale, i.e. $\tau_\T{I}=\tau_\T{II}=1$; (ii) Layer I updates faster, i.e. $\tau_\T{I}=1$ and $\tau_\T{II}>1$; and (iii) Layer II updates faster $\tau_\T{I}>1$ and $\tau_\T{II}=1$.

We seek full control of the system {as defined by Kalman~\cite{KAL60}}, meaning that with the proper choice of $\V u(t)$, we can steer the system from any initial state to any final state in finite time. To characterize controllability, we aim to design a matrix $\M B$ such that the system is controllable and the number of independent control inputs, $M$, is minimized. The minimum number of inputs, $N_\T i$, serves as our {measure of how difficult it is to control the system}.

To find a robust and efficient algorithm to determine $N_\T i$, we rely on the framework of structural controllability~\cite{LIN74}. We say that a matrix $\M A^*$ has the same structure as $\M A$, if the zero/nonzero elements of $\M A$ and $\M A^*$ are in the same position, and only the value of the nonzero entries can be different, in other words, in the corresponding network the links connect the same nodes, only the link weights can differ. A linear system of Eq.~(\ref{eq:dynamics}) defined by matrices $(\M A_\T{I}, \M A_\T{II}, \M D, \M B)$ is structurally controllable if there exists matrices with the same structure $(\M A_\T{I}^*, \M A_\T{II}^*, \M D^*, \M B^*)$ such that the dynamics defined by $(\M A_\T{I}^*, \M A_\T{II}^*, \M D^*, \M B^*)$ are controllable {according to the definition of Kalman}. Note that ultimately we are interested in controllability and not structural controllability. Yet, structural controllability is a useful tool because (i)~if a linear system is structurally controllable, it is controllable for almost all link weight combinations~\cite{LIN74} and (ii)~determining structural controllability can be mapped to a graph combinatorial problem allowing for efficient and numerically robust algorithms.

\section{Minimum input problem}\label{sec:min_input_prob}

Before addressing the minimum input problem of multiplex networks, we revisit the case of single-layer networks by providing an alternative explanation of the Minimum Input Theorem of Liu et al.~\cite{LIU11}. This new approach readily lends itself to be extended to multiplex, multi-timescale networks. Thus providing the basis for Sec.~\ref{sec:multiplex_net}, in which we develop an algorithm to determine $N_\T i$ for two-layer multiplex networks.

\subsection{Single-layer networks}\label{sec:single-layer_net}

\begin{figure}
	\centering
	\includegraphics[scale=.55]{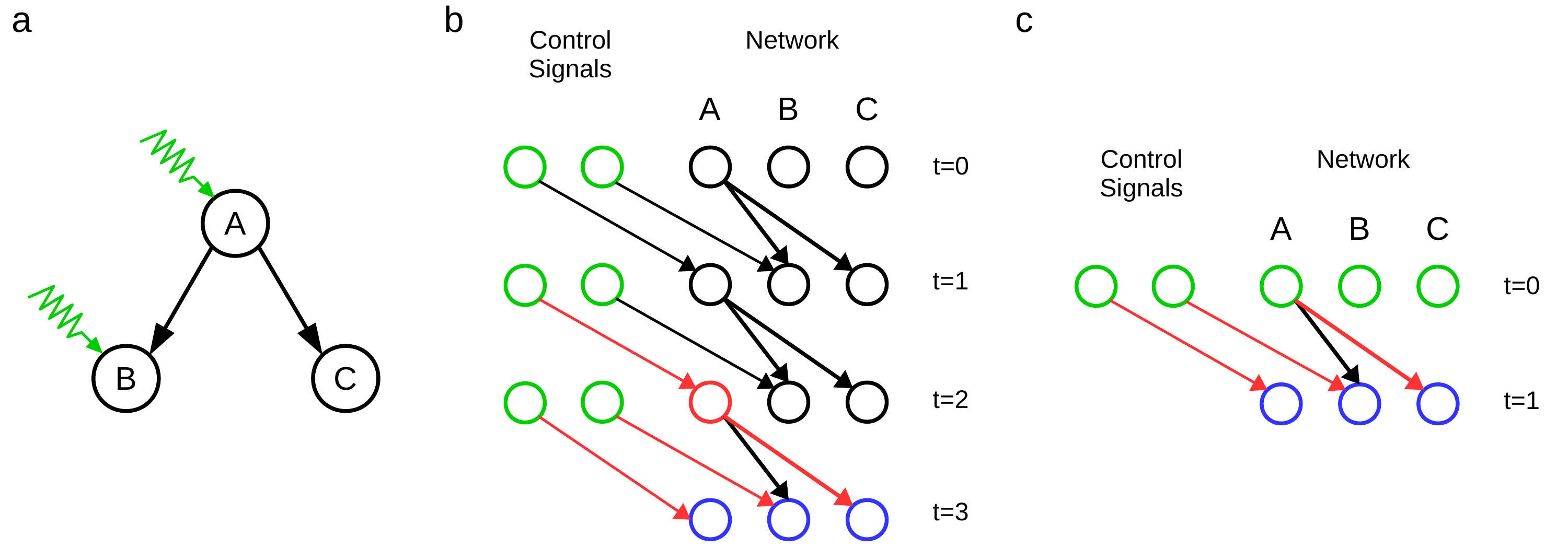}
	\caption{{\bf Structural controllability of single-layer networks.} {\bf (a)}~A single-layer network, we apply inputs to nodes $v_A$ and $v_B$. {\bf (b)}~The dynamic graph $\mathcal D_N$ representing the time evolution of the dynamics from $t=0$ to $t=N$. The system is controllable, because we can connect the set of nodes representing control signals (green) to the set of nodes at $t=N$ (blue) via disjoint paths (red). {\bf (c)}~The dynamic graph $\mathcal D_1$ representing the time evolution of the dynamics from $t=0$ to $t=1$. The system is controllable, because we can connect the control signals and nodes at $t=0$ (green) to the set of nodes at $t=1$ (blue) via disjoint paths (red), and all nodes are accessible from control signals.}
	\label{fig:singlenet}
\end{figure}

The linear discrete-time dynamics associated to a single-layer weighted directed network $\mathcal L$ are formulated as
\begin{equation}\label{eq:single-layer_dynamics}
\V{x}(t + 1) = \M{A} \V{x}(t) + \M{B}\V{u}(t),
\end{equation}
where $\V x(t)$, $\M A$, $\M B$ and $\V u(t)$ are defined similarly as in Eq.~(\ref{eq:dynamics}) (Fig.~\ref{fig:singlenet}a). To obtain a graph combinatorial condition for structural controllability we rely on the dynamic graph $\mathcal D_{T}$, which represents the time evolution of a system from $t=0$ to $t=T$~\cite{MUR87,PFI13,POS14b}. Each node $v_i$ in $\mathcal L$ is split into $T+1$ copies $\{v_{i,0},v_{i,1},\ldots,v_{i,T}\}$, each copy $v_{i,t}$ represents the state of node $v_i$ at time step $t$. We add a directed link $(v_{i,t} \rightarrow v_{j,t+1})$ for $t=0,1,\ldots,T-1$ if they are connected by a directed link $(v_{i}\rightarrow v_{j})$ in the original network, representing that the state of node $v_{j}$ at time $t+1$ depends on the state of its in-neighbours at the previous time step. To account for the controller, for each independent input we create $T$ nodes $u_{i,t}$ ($i=1,2,\ldots, M$; $t=0,1,\ldots,T-1$) each representing a control signal (i.e. the value of the $i$th input at time step $t$). We draw a directed link $(u_{i,t}\rightarrow v_{j,t+1})$ for $t=0,1,\ldots,T-1$ if $b_{ji}\neq 0$, where $b_{ji}$ is an element of matrix $\M B$.

According to Theorem 15.1 of Ref.~\cite{MUR87}, a linear system $(\M A, \M B)$ is structurally controllable if and only if in the associated dynamic graph $\mathcal D_N$ node sets $U=\{u_{i,t}\vert i=1,2,\ldots,M;t=0,1\ldots,N-1\}$ (green nodes in Fig.~\ref{fig:singlenet}b) and $V_N=\{v_{i,t=N}\vert i=1,2\ldots,N\}$ (blue nodes) are connected by $N$ disjoint paths (red links), i.e. there exists a set of disjoint paths $C=\{P_1,P_2,\ldots,P_N\}$ such that $U$ contains the set of starting points and $V_T$ is the set of endpoints. A path $P$ of length $l$ between node $v_{i_0}$ and $v_{i_l}$ is a sequence of $l$ consecutive links $[(v_{i_0}\rightarrow v_{i_1}),(v_{i_1}\rightarrow v_{i_2}),\ldots,(v_{i_{l-1}}\rightarrow v_{i_l})]$ such that each node is traversed only once. Node $v_{i_0}$ is the starting point and $v_{i_l}$ is the endpoint of $P$. Two paths $P_1$ and $P_2$ are disjoint if no node is traversed by both $P_1$ and $P_2$, a set of paths is disjoint if all paths in the set are pairwise disjoint.

A possible interpretation of this result is that if a $P_i$ path has starting point $u_{j,t_0}$ and endpoint $v_{k,t_1}$, we say that the signal $u_{j}(t_0)$ is assigned to set $x_k(t_1)$, the state of node $v_k$ at time $t_1$, through path $P_i$. Therefore we refer to path $P_i$ as a control path. The clear meaning of the dynamic graph and the control paths makes this condition useful to formulate proofs and to interpret results. However, it is rarely implemented to test controllability of large networks, because the size of the dynamical graph grows as $N^2$, rendering such algorithms too slow. In the following, we provide a condition that only requires the dynamic graph $\mathcal D_1$ as input; therefore it is more suitable for practical purposes.

It was shown in Refs.~\cite{MUR87,LIU11,COM13} that a linear system $(\M A, \M B)$ is structurally controllable if and only if (i)~in  $\mathcal D_1$ we can connect nodes $U\cup V_0=\{u_{i,t=0}\vert i=1,2,\ldots,M\}\cup\{v_{i,t=0}\vert i=1,2\ldots,N\}$ (green nodes in Fig.~\ref{fig:singlenet}c) and nodes $V_1=\{v_{i,t=1}\vert i=1,2\ldots,N\}$ (blue nodes) via $N$ disjoint paths (red links) and (ii)~all nodes are accessible from the inputs. This result can be understood as a self-consistent version of the previous condition involving $\mathcal D_N$: Instead of keeping track of the entire control paths as we previously did, we concentrate on a single time step. Consider the dynamic graph $\mathcal D_1$ representing the time evolution of the system from $t=0$ to $t=1$, and assume that the system is controllable. By definition we can set the state of each node independently at $t=0$; therefore we can treat them as control signals to control the system at a later time step. Now let us aim to control the system at $t=1$, according to our previous condition, it is necessary that $N$ disjoint paths exist between nodes $U\cup V_0=\{u_{i,t=0}\vert i=1,2,\ldots,M\}\cup\{v_{i,t=0}\vert i=1,2\ldots,N\}$ and nodes $V_1=\{v_{i,t=1}\vert i=1,2\ldots,N\}$. This is exactly requirement (i), together with the accessibility requirement (ii) it is a sufficient and necessary condition. Note that $D_1$ is a bipartite network (each link is connected to exactly one node in $U\cup V_0$ and one node in $V_1$) and each disjoint path in $\mathcal D_1$ is a single link.


The minimum input problem aims to identify the minimum number of inputs that guarantee controllability for a given network, in other words, the goal is to design a $\M B\in\mathbb{R}^{N\times M}$ for a given $\M A$ such that $M$ is minimized. For this we consider the dynamic graph $\mathcal D_1$ without nodes representing control signals. We then find a maximum cardinality matching, where a matching is a set of links that do not share an endpoint. The matching is a set of disjoint paths connecting node sets $V_0$ and $V_1$. Controllability requires $N$ disjoint paths between $U\cup V_0$ and $V_1$; therefore $N_\T i = N - N_\T{match}$, where $N_\T{match}$ is the size of the maximum matching (if $N_\T{match}=N$, $N_\T i=1$). Allowing the inputs to be connected to multiple nodes we can guarantee that all nodes are accessible from the inputs. Thus we recovered the Minimum Input Theorem of Liu et al.~\cite{LIU11}.

In summary, by relying on a self-consistent condition for structural controllability we re-derived the known result that identifying $N_\T i$ is equivalent to finding a maximum matching in $\mathcal D_1$. In the next section we show that this new self-consistent approach lends itself to be extended to the multiplex, multi-timescale model defined by Eq.~(\ref{eq:dynamics}), allowing us to derive analogous method to identify $N_\T i$. 

\subsection{Multiplex networks}\label{sec:multiplex_net}

To find the minimum number of inputs $N_\T i$ for multiplex, multi-timescale networks, we first extend the definition of the dynamic graph. We define the dynamic graph $\mathcal D_{\tau_\T{II}}$ such that it captures the time evolution of a multiplex system defined by $(\M A_\T{I}, \M A_\T{II}, \M D, \M B)$ and Eq.~(\ref{eq:dynamics}) from $t=0$ to $t=\tau_\T{II}$. For sake of brevity, we assume that $\tau_\T{I} = 1$ and $\tau_\T{II} \geq 1$, the case of $\tau_\T{I} > 1$ and $\tau_\T{II} = 1$ is treated similarly (Fig.~\ref{fig:3examples}d). Each node $v_i^\T{I}$ in Layer I is split into $\tau_\T{II}+1$ copies $\{v^\T{I}_{i,0},v^\T{I}_{i,1},\ldots,v^\T{I}_{i,\tau_\T{II}}\}$; each node $v_i^\T{II}$ in Layer II is split into two copies $\{v^\T{II}_{i,0},v^\T{II}_{i,\tau_\T{II}}\}$, because Layer II does not update during the intermediate time steps. We draw a link from $v^\T{I}_{i,t}$ to $v^\T{I}_{j,t+1}$ ($t=0,1,\ldots,\tau_\T{II}-1$) if they are connected in Layer I by a directed link $(v^\T{I}_{i}\rightarrow v^\T{I}_{j})$, and similarly we connect $v^\T{II}_{i,0}$ to $v^\T{II}_{j,\tau_\T{II}}$ if they are connected in Layer II. In addition we draw a link between each pair $v^\T{I}_{i,0}$ and $v^\T{II}_{i,\tau_\T{II}}$ to account for the interconnectedness.

As a natural extension of self-consistent approach introduced in Sec.~\ref{sec:single-layer_net}, assume that the system is controllable. If the system is controllable, we can set the state of each node independently at $t=0$. To control the system at $t=\tau_\T{II}$, all nodes at $t=\tau_\T{II}$ in  $\mathcal D_{\tau_\T{II}}$ (blue nodes in Fig.~\ref{fig:3examples}) have to be connected to a node at $t=0$ or to a control signal (green nodes) via a disjoint path (red links). In other words, a linear two-layer system $(\M A_\T{I}, \M A_\T{II}, \M D, \M B)$ is structurally controllable only if there exists $2N$ disjoint paths in the dynamic graph connecting node set $U\cup V_0 =\{u_{i,t}\vert i=1,2,\ldots,M;t=0,1,\ldots,\tau_\T{II}-1\}\cup \{v^\T{I}_{i,0}\vert i=1,2,\ldots,N\}\cup\{v^\T{II}_{i,0}\vert i=1,2,\ldots,N\}$ and node set $V_{\tau_\T{II}}=\{v^\T{I}_{i,\tau_\T{II}}\vert i=1,2,\ldots,N\}\cup\{v^\T{II}_{i,\tau_\T{II}}\vert i=1,2,\ldots,N\}$. In other words, a linear two-layer system $(\M A_\T{I}, \M A_\T{II}, \M D, \M B)$ is structurally controllable only if there exists $2N$ disjoint paths in the dynamic graph connecting node set $U\cup V_0=\{u_{i,t}\vert i=1,2,\ldots,M;t=0,1,\ldots,\tau_\T{II}-1\}\cup \{v^\T{I}_{i,0}\vert i=1,2,\ldots,N\}\cup\{v^\T{II}_{i,0}\vert i=1,2,\ldots,N\}$ and node set $V_{\tau_\T{II}}=\{v^\T{I}_{i,\tau_\T{II}}\vert i=1,2,\ldots,N\}\cup\{v^\T{II}_{i,\tau_\T{II}}\vert i=1,2,\ldots,N\}$. 

To test whether the system is controllable by $M$ independent inputs, we need to find a $\M B\in\mathbb R^{N\times M}$ such that the system is controllable. We do not have to check all possibilities, because if such $\M B$ exists, then the system is also controllable for $\M B^\prime\in\mathbb R^{N\times M}$ where $\M B^\prime$ has no zero elements; therefore, we only check the case when each input is connected to each node in Layer I. Given matrices $(\M A_\T{I}, \M A_\T{II}, \M D, \M B^\prime)$, we now have to count the number of disjoint paths connecting $U\cup V_0$ and $V_{\tau_\T{II}}$ in the corresponding dynamic graph $\mathcal D_\T{II}$. We find these paths using maximum flow: We set the capacity of each link and each node to 1, we then find the maximum flow connecting source node set $U\cup V_0$ to target node set $V_{\tau_\T{II}}$ using any maximum flow algorithm of choice. If the system is structurally controllable, the maximum flow equals to $2N$; if it is less than $2N$, additional inputs are needed.

 We can now identify the minimum number of inputs $N_\T i$ by systematically scanning possible values of $M$. A simple approach is to first set $M=1$, and test if the system is controllable. If not, increase $M$ by one. Repeat this until the smallest $M$ yielding full control is found. Significant increase in speed is possible if we find the minimum value of $M$ using bisection. We initially know that $N_\T{i}^\T{upper}=N \geq N_\T i \geq N_\T{i}^\T{lower}=1$. We set $M=(N_\T{i}^\T{upper}+N_\T{i}^\T{lower})/2$, and test if the system is controllable. If yes, we set $N_\T{i}^\T{upper}=M$; if no, we set $N_\T{i}^\T{lower}=M$. We repeat this until $N_\T{i}^\T{upper}=N_\T{i}^\T{lower}$, which provides $N_\T i$. For implementation, we used Google OR-tools and igraph python packages~\cite{CSA06,ORTOOLS}.

The one-to-one coupling between Layer I and Layer II guarantees that full control is possible with at most $N$ independent inputs; therefore we often normalize $N_\T i$ by $N$, i.e. $n_\T i = N_\T i / N$.

Note that in the above argument we rely on the test of structural controllability based on the dynamic graph, which was originally introduced for single-timescale networks~\cite{MUR87}. The sufficiency of the condition relies on the fact that the zero is the only degenerate eigenvalue of a matrix $\M A$ if the nonzero elements of $\M A$ are uncorrelated. However, this might not remain true for the spectrum of $\M A^\tau$, where $\tau>1$, due to correlations arising in the nonzero elements of $\M A^\tau$. If a $\lambda\neq 0$ eigenvalue has larger geometric multiplicity than the multiplicity of $0$, $N_\T i$ would be larger than predicted by the dynamic graph; if a  $\lambda\neq 0$ eigenvalue has larger geometric multiplicity than $1$ but smaller than the multiplicity of zero, it does not affect $N_\T i$, but may require connecting an input to multiple nodes~\cite{YUA13}. In the $\tau_\T{I}>0$ and $\tau_\T{II}=1$ case, a control signal is only injected into Layer II every $\tau_\T{I}$ time step (Fig.~\ref{fig:3examples}d); therefore, the spectrum of $\M A_\T{II}^{\tau_\T{I}}$ becomes relevant. However, we are interested in large and sparse complex networks whose spectra is dominated by the zero eigenvalue~\cite{YUA13}. Therefore it is reasonable to expect that the spectrum of $\M A^\tau$ will be dominated by zero eigenvalues as well. Meaning that the minimum number of inputs is correctly given by this graph combinatorial condition. Furthermore the one-to-one coupling between the layers guarantees that control is possible by only interacting with Layer I directly. 

So far, we developed a method to characterize controllability of a {multiplex}, multi-timescale system based on the underlying network structure and the timescale of each of its layers. In the next section, we rely on these tools to systematic study how network characteristics and timescales affect $N_\T i$.

\section{Results}\label{sec:results}

In this section we investigate how different timescales and the degree distribution of each layer affect controllability. For timescales, we consider three scenarios: (i) the subsystems operate on the same timescale, i.e. $\tau_\T{I}=\tau_\T{II}=1$; (ii) Layer I updates faster, i.e. $\tau_\T{I}=1$ and $\tau_\T{II}>1$; and (iii) Layer II updates faster $\tau_\T{I}>1$ and $\tau_\T{II}=1$. To uncover the effect of degree distribution, we consider layers with Poisson (ER) or scale-free (SF) degree distribution, the latter meaning that the distribution has a power-law tail.

We generate scale-free layers using the static model~\cite{GOH01}: We start with $N$ unconnected nodes. Each node $v_i$ is assigned two hidden parameters $w_\text{in}(i)=i^{-\zeta_\text{out}}$ and $w_\text{out}(i)=i^{-\zeta_\text{out}}$, where $i=1,2,\ldots,N$. The weights are then shuffled to eliminate any correlations of the in- and out-degree of individual nodes and between layers. We then randomly place $L$ directed links by choosing the start- and endpoint of the link with probability proportional to $w_\text{in}(i)$ and $w_\text{out}(i)$, respectively. For large $N$ this yields the degree distribution
\begin{equation}
P^{\text{SF}}_\text{in/out}(k) = \frac{\left[c(1-\zeta_\text{in/out})^{1/\zeta_\text{in/out}}\right]}{\zeta_\text{in/out}}\frac{\Gamma(k-1/\zeta_\text{in/out},c[1-\zeta_\text{in/out}])}{\Gamma(k+1)},
\end{equation}
where $c=L/N$ is equal to the average degree, and $\Gamma(n,x)$ is the upper incomplete gamma function. For large $k$, $P_\text{in/out}^{\text{SF}}(k)\sim k^{-(1+1/\zeta_\text{in/out})} = k^{-\gamma_\text{in/out}}$, where $\gamma_\text{in/out}=1+1/\zeta_\text{in/out}$ is the exponent characterizing the tail of the distribution.

To reduce the number of parameters we only study layers with symmetric degree distribution, e.g. $P(k)=P_\T{in}(k)=P_\T{out}(k)$; however, the in- and out-degree of a specific node can be different.

\subsection{No timescale separation (\textnormal{$\tau_\T{I}=\tau_\T{II}=1$})}
\label{sec:noseparation}

In the special case when both layers operate on the same timescale, i.e. $\tau_\T{I}=\tau_\T{II}=1$ (Fig.~\ref{fig:3examples}b), there is no qualitative difference between the dynamics of the layers. The reason why the system cannot be treated as a single large network is that we are only allowed to directly interact with Layer I. Recently Iudice et al. developed methodology to identify $N_\T i$ if the control signals can only be connected to a subset of nodes~\cite{IUD15}. However, the one-to-one coupling between the layers enables us to find $N_\T i$ using a simpler approach.

Finding $N_\T i$ for a single-layer network is equivalent to finding a maximum matching of the network~\cite{LIU11}. A matching is a set of directed links that do not share starting or end points, and a node is unmatched if there is no link in the matching pointing at it. Liu et al. showed that full control of a network is possible if each unmatched node is controlled directly by an independent input; therefore $N_\T i$ is provided by the minimum number of unmatched nodes. To determine $N_\T i$ for a two-layer network, we first find a maximum matching of the combined network of Layer I and Layer II. If there are no unmatched nodes in Layer II, we only have to interact with Layer I; therefore we are done. If a node $v^\T{II}_i$ is unmatched in Layer II,  $v^\T{I}_i$ is necessarily matched by some node $v^\T{I}_j$, otherwise the size of the matching could be increased by adding $(v_i^\T{I}\rightarrow v_i^\T{II})$. By taking out the link $(v^\T{I}_j\rightarrow v^\T{I}_i)$ from the matching and including $(v^\T{I}_i\rightarrow v^\T{II}_i)$ the size of the maximum matching does not change, and we moved the unmatched node from Layer II to Layer I. We repeat this for all unmatched nodes in Layer II. (Note that it may be necessary to connect inputs to additional nodes so that all nodes are reached by the control signals. Due to the one-to-one coupling between the layers this too can be accomplished by interacting only with Layer I.) This simplified method allows faster identification of $N_\T i$ using the Hopcroft-Karp algorithm~\cite{HOP73} and analytically solving $n_\T i=N_\T i /N$ for random networks based on calculating the fraction of always matched nodes as described in Appendix~\ref{app:sec:analytical}~\cite{ZDE06,JIA13,JIA14,POS14}.

\begin{figure}
	\centering
	\includegraphics[scale=.55]{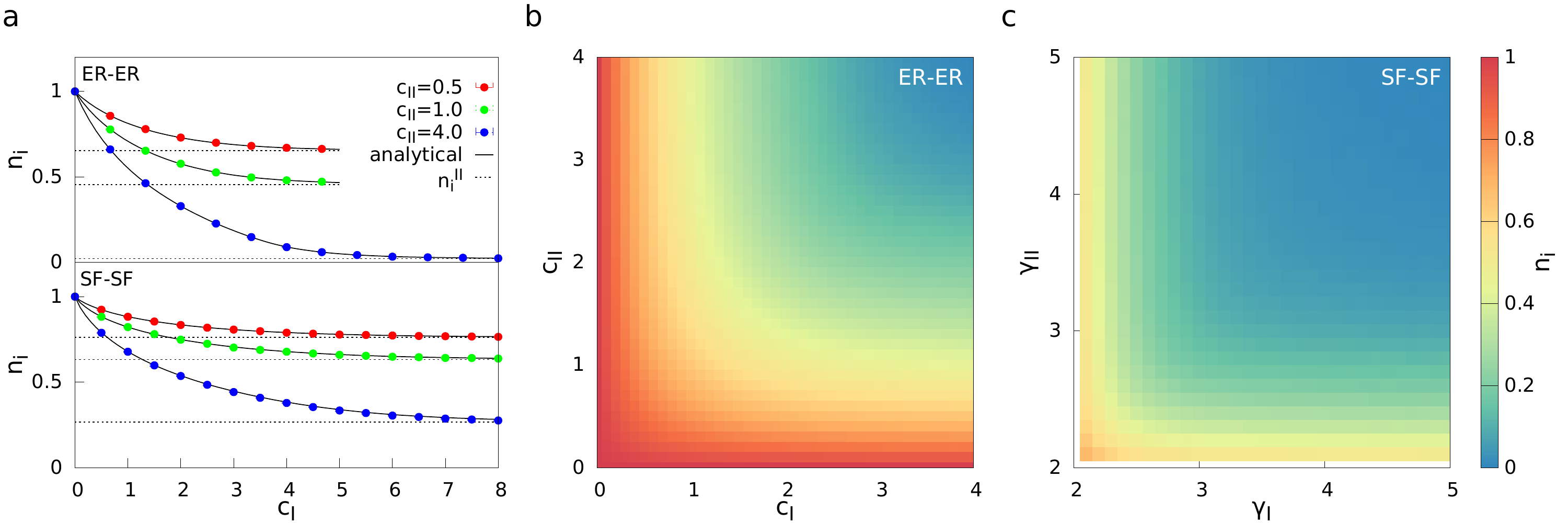}
	\caption{{\bf No timescale separation.} {\bf (a)}~Number of inputs $n_\T i$ in function of $c_\T{I}$ for ER-ER and SF-SF ($\gamma_\T{I}=\gamma_\T{II}=2.5$) networks. The circles represent simulations, the continuous line is the analytical solution, and the dashed line is the analytical solution of $n^\T{II}_\T i$, the number of independent inputs necessary to control Layer II in isolation~\cite{LIU11}. {\bf (b)}~$n_\T i$ for ER-ER networks with varying average degrees $c_\T{I}$ and $c_\T{II}$. In both layers $P(k)=P(k_\T{in})=P(k_\T{out})$, therefore the heatmap is symmetric with respect to the diagonal. Increasing $c$ in either layer enhances controllability. {\bf (c)}~$n_\T i$ for SF-SF networks with $c_\T{I}=c_\T{II}=4.0$ and varying degree exponents $\gamma_\T{I}$ and $\gamma_\T{II}$. Increasing degree heterogeneity in either layer  increases $n_\T i$. Each data point is the average over $10$ randomly generated networks with $N=10,000$. The standard deviation of the measurements remains below 0.01.}
	\label{fig:secIIIA}
\end{figure}

First, we measure $n_\T i$ while fixing the average degree of Layer II ($c_\T{II}$) and varying the average degree of Layer I ($c_\T{I}$). For both ER-ER and SF-SF networks, we find that $n_\T i$ decreases for increasing values of $c_\T{I}$ and converges to $n_\T i^\T{II}=N_\T i^\T{II}/N$, the normalized number of inputs needed to control Layer II in isolation~(Fig.~\ref{fig:secIIIA}a). The latter observation is easily understood: $n_\T i$ is determined by the fraction of unmatched nodes in the combined network of the two layers; if $c_\T{I}$ is high enough, Layer I is perfectly matched; therefore all unmatched nodes are in Layer II. Based on the same argument, $n_\T i^\T{I}$ also serves as a lower bound for $n_\T i$.

Varying both $c_\T{I}$ and $c_\T{II}$ for ER-ER and both $\gamma_\T{I}$ and $\gamma_\T{II}$ for SF-SF with constant average degrees $c_\T{I}=c_\T{II}$, we find that dense networks require less inputs than sparse networks (Fig.~\ref{fig:secIIIA}b) and degree heterogeneity makes control increasingly difficult (Fig.~\ref{fig:secIIIA}c) -- in line with results for single-layer networks~\cite{LIU11}. We also observe that $n_\T i$ is invariant to exchanging Layer I and Layer II. This is explained by the fact that the size of the maximum matching is invariant to flipping the direction of all links, and on the ensemble level this is the same as swapping the two layers for networks with $P(k_\T{in})=P(k_\T{out})$.

In summary, for no timescale separation controllability is equally affected by the network structure of both layers, and $n_\T i$ is greater or equal to the number of inputs necessary to control any of its layers in isolation. Similarly to single-layer networks, networks with low average degree and high degree heterogeneity require more independent inputs than sparse homogeneous networks.

\subsection{Layer I updates faster (\textnormal{$\tau_\T{I}=1$, $\tau_\T{II}>1$})}

In the previous section we found that the network structure of the two layers equally affect $n_\T i$ if $\tau_\T{I}=\tau_\T{II}=1$. This is not the case if the timescales are different, for example if Layer I updates faster than Layer II, we expect that we need fewer inputs than in the same timescale case by the virtue of having more opportunity to interact with the faster system (Fig.~\ref{fig:3examples}b). In this section we systematically study this effect using the algorithm described in Sec.~\ref{sec:multiplex_net} and analytical arguments.

\begin{figure}
	\centering
	\includegraphics[scale=.55]{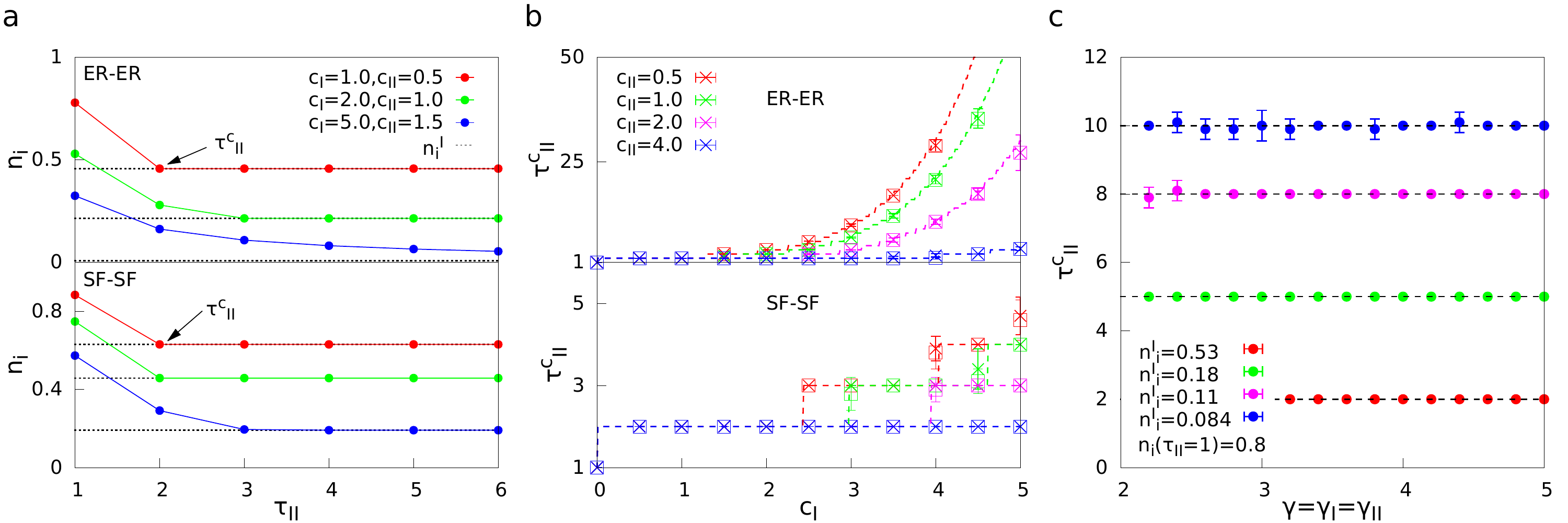}
	\caption{{\bf Layer I updates faster.} {\bf (a)}~Number of inputs $n_\T i$ for single ER-ER and SF-SF ($\gamma_\T{I}=\gamma_\T{II}=2.5$) networks  with $N=10,000$ and varying timescale parameter $\tau_\T{II}$. The number of inputs $n_\T i$ monotonically decreases with increasing $\tau_\T{II}$, and for $\tau_\T{II}\geq \tau_\T{II}^\T c$, $n_\T i=n_\T i^\T{I}$. {\bf (b)}~The critical timescale parameter $\tau_\T{II}^\T c$ for ER-ER and SF-SF ($\gamma_\T{I}=\gamma_\T{II}=2.5$) networks with varying average degree $c_\T{I}$ and $c_\T{II}$. The crosses represent direct measurements of $\tau_\T{II}^\T c$; the squares represent the approximation obtained by applying Eq.~(\ref{eq:tauIIc_approx}) to measurements of $n_\T i(\tau_\T{II}=1)$ and $n^\T{I}_\T i$; and the dashed line is an approximation obtained using analytically calculated expectation values of $n_\T i(\tau_\T{II}=1)$ and $n^\T{I}_\T i$. {{\bf (c)}~We measure $\tau_\T{II}^\T c$ for SF-SF networks with the same $n_\T i(\tau_\T{II}=1)$ and $n^\T I_\T i$ as a function of $\gamma=\gamma_\T{I}=\gamma_\T{II}$. Equation~(\ref{eq:tauIIc_approx}) predicts that $\tau_\T{II}^\T c$ remains constant (dashed line), in line with our observations.} For (b-c), each data point is the average over $10$ randomly generated networks with $N=10,000$ and error bars represent the standard deviation.}	
	\label{fig:secIIIB}
\end{figure}

By measuring $n_\T i$ for ER-ER and SF-SF networks as a function of $\tau_\T{II}$, we find that $n_\T i$ monotonically decreases with increasing $\tau_\T{II}$ (Fig.~\ref{fig:secIIIB}a), confirming our expectations. For both ER-ER and SF-SF networks $n_\T i(\tau_\T{II})$ converges to $n_\T i^\T{I}=N_\T i^\T{I}/N$ which is the normalized number of inputs needed to control Layer I in isolation. This can be understood by the following argument: Suppose that $\tau_\T{II}=N$, the maximum number of time steps needed to impose control on any network with $N$ nodes~\cite{KAL63}. We use the state of Layer I at $t=0$ to set the state of Layer II at $t=N$, and we have $N$ time steps to impose control on Layer I as if it was just by itself. For a given network we define the critical timescale parameter $\tau^\T c_\T{II}$ as the minimum value of $\tau_\T{II}$ for which $n_\T i(\tau_\T{II})=n_\T i^\T{I}$. Above the critical timescale separation, Layer I completely determines $n_\T i(\tau_\T{II})$ independent of the structure of Layer II, in other words, the {multiplex} nature of the system no longer plays a role in determining $n_\T i$. 

Measuring $\tau_\T{II}^\T c$ we find that for both ER-ER and SF-SF networks $\tau_\T{II}^\T c$ monotonically increases with increasing $c_\T I$ for fixed $c_\T{II}$, and decreases with increasing $c_\T{II}$ for fixed $c_\T{I}$ (Fig~\ref{fig:secIIIB}b). That is $\tau_\T{II}^\T c$ is the highest if Layer I is dense and Layer II is sparse. SF-SF networks have significantly lower  $\tau_\T{II}^\T c$ than ER-ER networks with the same average degree.

To understand the observed pattern we provide an approximation to calculate $\tau_\T{II}^\T c$. We call a node $v^\T{I}_i$ externally controlled if in the dynamic graph $v^\T{I}_{i,\tau_\T{II}}$ is connected to an external signal $u_{j,t}$ via a disjoint control path (e.g. nodes $v^\T{I}_A$ and $v^\T{I}_B$ in Fig.~\ref{fig:3examples}c), and  the number of such nodes is denoted by $N_\T e(\tau_\T{II})$. We have previously shown that we require $N_\T i^\T I$ independent inputs at $\tau_\T{II}^\T c$. For each independent input and each time step, we have one control signal $u_{i,t}$; therefore we need timescale parameter
\begin{equation}
\label{eq:tauIIc}
\tau_\T{II}^\T c=\lceil N_\T e(\tau_\T{II}^\T c)/N_\T i^\T I\rceil
\end{equation}
to insert enough signals required by the $N_\T e(\tau_\T{II}^\T c)$ externally controlled nodes, where $\lceil\cdot\rceil$ is the ceiling function.
Equation (\ref{eq:tauIIc}) is not yet useful as it contains $\tau_\T{II}^\T c$ on both side. Observing that $N_\T e(\tau_\T{II})$ is a monotonically increasing function of $\tau_\T{II}$ and $N_\T e(\tau_\T{II}=1)= N_\T i(\tau_\T{II}=1)$, we can write
\begin{equation}
N_\T i(\tau_\T{II}=1) \leq N_\T e(\tau_\T{II}^\T c)\leq N.
\end{equation}
In the special case when Layer II is fully connected, $\tau_\T{II}^\T c=1$ and $N_\T e(\tau_\T{II}^\T c=1)= N_\T i(\tau_\T{II}=1)$. In the case when Layer II is entirely disconnected, i.e. is composed of isolated nodes, $N_\T e(\tau_\T{II})=N=N_\T i(\tau_\T{II}=1)$. These two opposite limiting cases suggest that it is reasonable to approximate $N_\T e(\tau_\T{II}^\T c)$ by its lower bound:
\begin{equation}
\label{eq:tauIIc_approx}
\tau_\T{II}^\T c\approx\lceil N_\T i(\tau_\T{II}=1)/N_\T i^\T I\rceil,
\end{equation}
which entirely depends on quantities that we can easily measure or analytically compute. We find that Eq.~(\ref{eq:tauIIc_approx}) preforms remarkably well: Figure~\ref{fig:secIIIB}b compares direct measurements of $\tau_\T{II}^\T c$ to approximations obtained by using measurements and analytically computed values of $N_\T i(\tau_\T{II}=1)$ and $N_\T i^\T I$. The approximation based on measurements out performs the analytical calculations, because the analytical results provide the expectation value of the numerator and denominator for ER and SF network ensembles{;} and therefore the ceiling function is applied to the fraction of averages, instead of averaging after applying the ceiling function. {To further test the Eq.~(\ref{eq:tauIIc_approx}), we fix $n_\T i(\tau_\T{II}=1)$ and $n^\T I_\T i$ and we analytically calculate $c_\T{I}$ and $c_\T{II}$ for SF-SF networks with varying degree exponent $\gamma=\gamma_\T I=\gamma_\T{II}$ using the framework developed in Appendix~\ref{app:sec:analytical}. Then we generate SF-SF networks and measure $\tau_\T{II}^\T c$ as a function of $\gamma$. The approximation predicts that $\tau_\T{II}^\T c$ remains constant, in line with our observations~(Fig.~\ref{fig:secIIIB}c).}

The good performance of Eq.~(\ref{eq:tauIIc_approx}) is partly due to the role of the ceiling function, as it is insensitive to changes in the numerator that are small compared to $N_\T i^\T I$. Indeed, errors are more pronounced if $N_\T i(\tau_\T{II}=1)/N_\T i^\T I$ is close to an integer (e.g. data point $c_\T{I}=4.5$ and $c_\T{II}=1$ in Fig.~\ref{fig:secIIIB}b for ER-ER), or $N_\T i(\tau_\T{II}=1)\gg N_\T i^\T I$ (e.g. data points $n_\T i^\T I=0.084$ in Fig.~\ref{fig:secIIIB}c). 

What we learn from this approximation is that $\tau_\T{II}^\T c$ depends only indirectly on the degree distribution of Layer I and Layer II through the control properties of the system without timescale separation -- $N_\T i(\tau_\T{II}=1)$ and $N^\T I_\T i$. In Sec.~\ref{sec:noseparation}, we showed that $N_\T i(\tau_\T{II}=1)\geq N^\T{II}_\T i$, therefore $\tau_\T{II}^\T c$ is expected to be large if Layer I is easy to control (e.g. it is dense and has homogeneous degree distribution) and Layer II is hard to control (e.g. it is sparse and has heterogeneous degree distribution).

In summary, if Layer I updates faster, timescale separation enhances controllability up to a critical timescale parameter $\tau_\T{II}^\T c$, above which $n_\T i(\tau_\T{II})=n^\T I_\T i$  and is completely determined by Layer I. The critical timescale parameter $\tau_\T{II}^\T c$ largely depends on the controllability of the system without timescale separation, it is expected to be large if Layer I is easy and Layer II is hard to control.

\subsection{Layer II updates faster (\textnormal{$\tau_\T{I}>1$, $\tau_\T{II}=1$})}

\begin{figure}
	\centering
	\includegraphics[scale=.55]{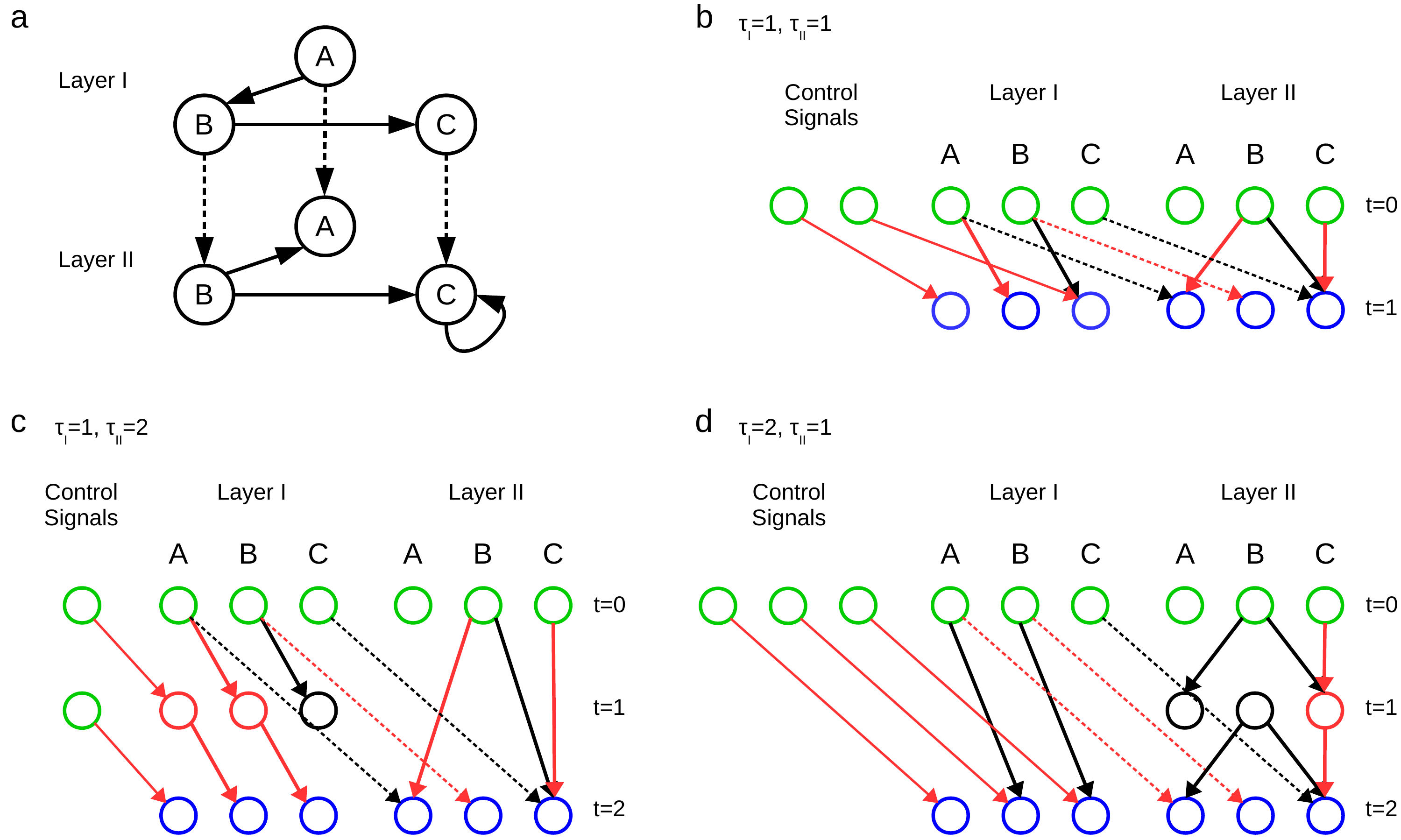}
	\caption{{\bf Structural controllability of two-layer {multiplex} networks.} {\bf (a)}~A two-layer network. {\bf (b-c)}~To determine $N_\T i$, we construct the dynamic graph representing the time evolution of the system from $t_0=0$ to $t_1=\max(\tau_\T{I},\tau_\T{II})$. The system is controllable {only} if all nodes at $t_1$ (blue) are connected to nodes at $t_0$ or nodes representing control signals (green) via disjoint paths (red). {\bf (b)}~In case of no timescale separation ($\tau_\T{I}=\tau_\T{II}=1$), each disjoint control path consists of a single link, yielding $N_\T i=2$. {\bf (c)}~If Layer I updates twice as frequently as Layer II ($\tau_\T{I}=1$, $\tau_\T{II}=2$), we are allowed to inject control signals at time steps $t=0$ and $1$, reducing the number of inputs to $N_\T i=1$.  {\bf (d)}~On the other hand, if Layer II is faster ($\tau_\T{I}=2$, $\tau_\T{II}=1$), Layer II needs to support longer control paths, yielding $N_\T i = 3$.}
	\label{fig:3examples}
\end{figure}

\begin{figure}
	\centering
	\includegraphics[scale=.55]{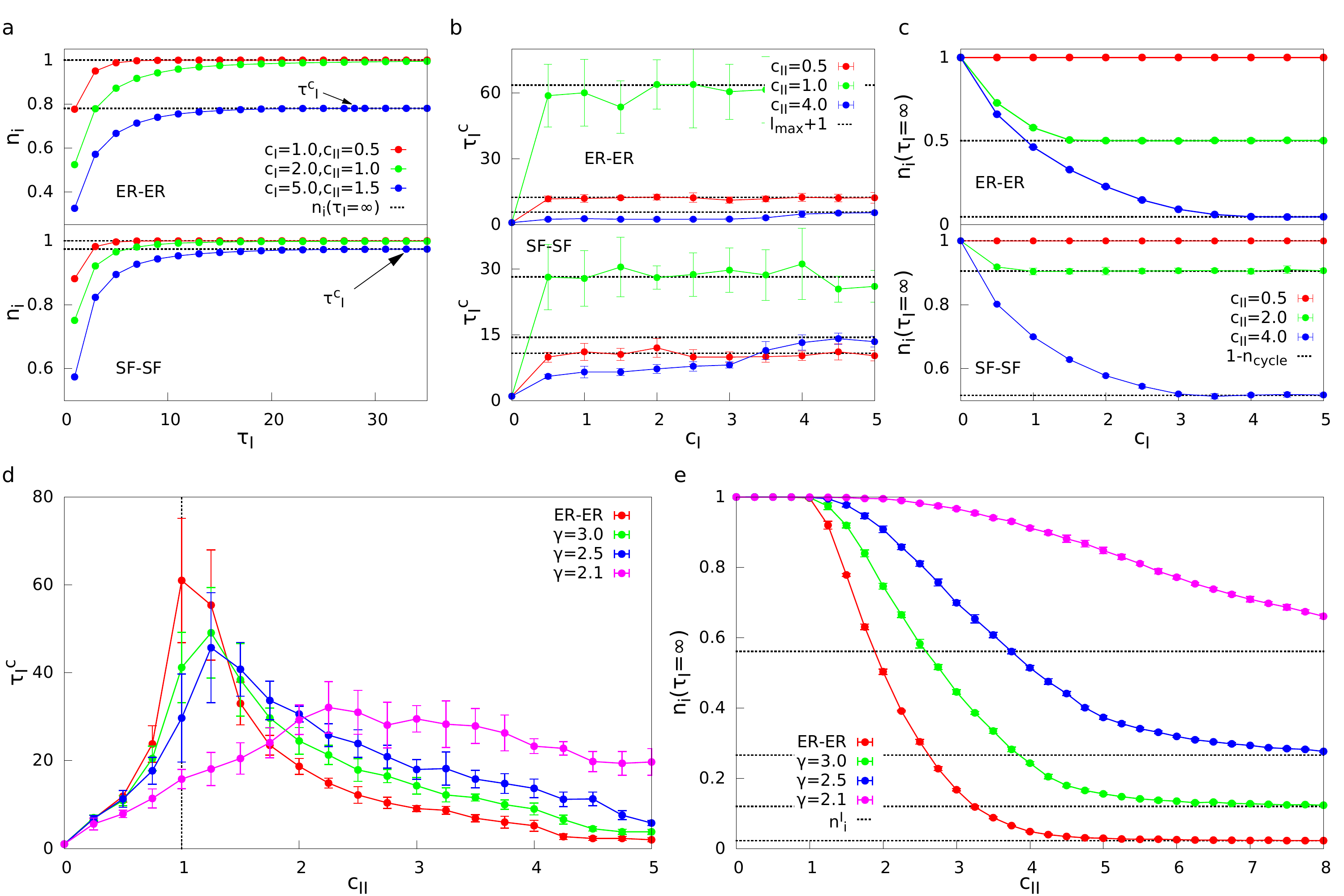}
	\caption{{\bf Layer {II} updates faster.} {\bf (a)}~Number of inputs $n_\T{i}$ for single ER-ER and SF-SF ($\gamma_\T{I}=\gamma_\T{II}=2.5$) networks with $N=10,000$ and varying timescale parameter $\tau_\T{I}$. The number of inputs $n_\T i$ monotonically increases with increasing $\tau_\T{I}$; for $\tau_\T{I}\geq \tau_\T{I}^\T c$, $n_\T i=n_\T i(\tau_\T{I}=\infty)$. {\bf (b)}~$\tau_\T{I}^\T c$ as a function of $c_\T I$. For $c_\T I\leq 1$,  $\tau_\T{I}^\T c$ quickly reaches its upper bound; for $c_\T I>1$, the convergence is somewhat delayed. {\bf (c)}~$n_\T{i}(\tau_\T{I}=\infty)$ as a function of $c_\T I$. Increasing $c_\T I$ facilitates control, until $n_\T i$ reaches its lower bound. {\bf (d)}~$\tau_\T{I}^\T c$ as a function of $c_\T{II}$ with fix $c_\T{I}=4.0$. The peak of $\tau_\T{I}^\T c$ corresponds to the critical point where the giant strongly connected component in Layer II emerges. {\bf (e)}~$n_\T{i}(\tau_\T{I}=\infty)$ as a function of $c_\T{II}$ with fix $c_\T{I}=4.0$. For $c_\T{II}<1$, Layer II does not contain cycles, therefore $n_\T i(\tau_\T{I}=\infty)=1$; for large $c_\T{II}$, Layer II can be completely covered with cycles, and $n_\T i(\tau_\T{I}=\infty)$ is determined by $n_\T i^\T I$. For (b-e), each data point is the average over $10$ randomly generated networks with $N=10,000$ and error bars represent the standard deviation.}	
	\label{fig:secIIIC}
\end{figure}

Finally we investigate the case when Layer II operates faster than Layer I, i.e. $\tau_\T{I}>1$ and $\tau_\T{II}=1$ (Fig.~\ref{fig:3examples}d). Measurements show that $n_\T i$ monotonically increases in function of $\tau_\T{I}$ for both ER-ER and SF-SF networks, and $n_\T i$ remains constant if $\tau_\T{I}\geq \tau^\T c_\T{I}$, where  $\tau^\T c_\T{I}$ is defined for a single network (Fig.~\ref{fig:secIIIC}a). To understand these results consider the following argument: Some nodes of Layer II are internally controlled, meaning that the state of these nodes at $t=\tau_\T{I}$ is set by the state of nodes within Layer II at $t=0$ connected to them via disjoint control paths (node $v^\T{II}_C$ in Fig.~\ref{fig:3examples}d); while the rest of the nodes of Layer II have to be controlled by nodes of Layer I. The maximum number of internally controlled nodes is set by the number of disjoint paths of length $\tau_\T{I}$. A directed open path traversing $l$ links in Layer II yields a path in the dynamic graph of at most length $l$; therefore if $\tau_\T I>l$ the path can no longer be used for control. For example, in Fig.~\ref{fig:3examples}a path $(v_B^\T{II}\rightarrow v_A^\T{II})$ consists of a single link; therefore, we can use it for control if $\tau_\T{I}=1$ (Fig.~\ref{fig:3examples}b) and it is no longer useful if $\tau_\T{I}>1$ (Fig.~\ref{fig:3examples}d). However, a cycle can support a path in the dynamic graph of any length, e.g. the self-loop $(v_C^\T{II}\rightarrow v_C^\T{II})$ in Fig.~\ref{fig:3examples}. This predicts that
\begin{equation}
\label{eq:tauI_ni_upperbound}
n_\T i(\tau_\T I=\infty)\geq 1-n_\T{cycle},
\end{equation}
where $n_\T{cycle}=N_\T{cycle}/N$ is the maximum fraction of nodes that can be covered with cycles in Layer II. Furthermore, it also means that
\begin{equation}
\label{eq:tauIc_upperbound}
\tau_\T{I}^\T{c}\leq l_\T{max}+1,
\end{equation}
where $l_\T{max}$ is the maximum length of a control path that does not involve cycles, a quantity that only depends on the structure of Layer II. We provide the formal definition $l_\T{max}$ and algorithms to measure $n_\T{cycle}$ and $l_\T{max}$ in Appendix~\ref{app:sec:algorithms}.

\begin{figure}
	\centering
	\includegraphics[scale=.55]{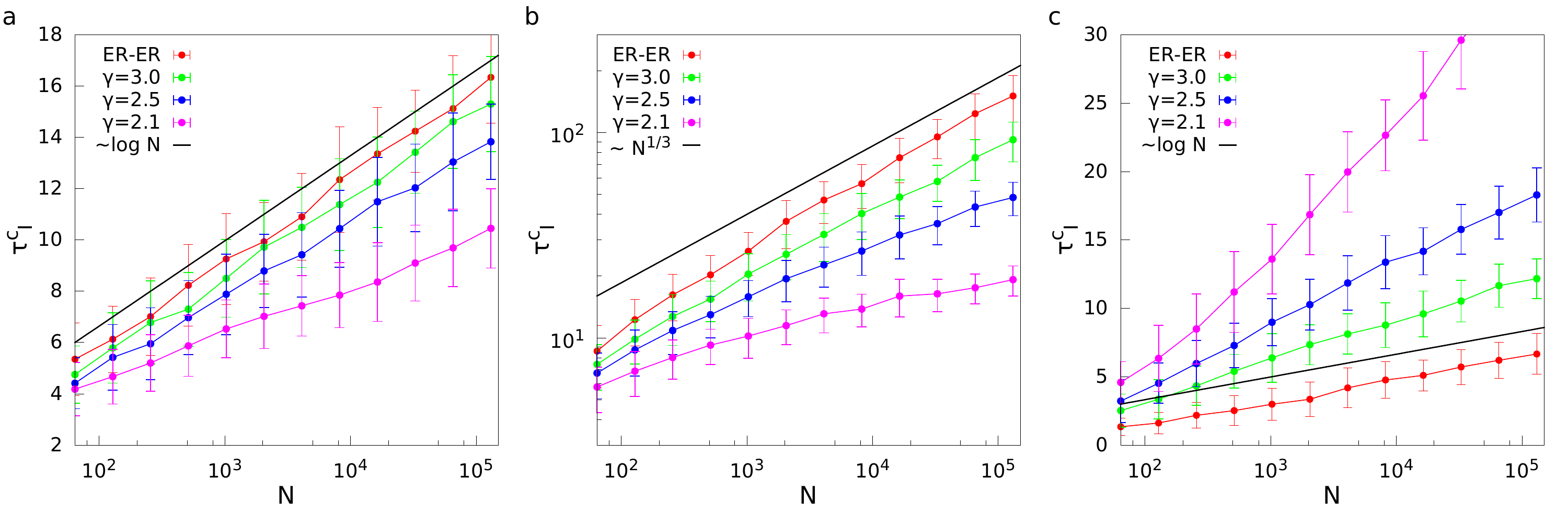}
	\caption{{\bf Layer II updates faster -- Network size effects.} Critical timescale parameter for ER-ER networks and SF-SF networks with varying network size $N$. {\bf (a)}~Layer II has no giant strongly connected component ($c_\T{II}=0.5<1$), $l_\T{max}$ equals the diameter $D$ of Layer II which scales as $D\sim \log N$ for ER networks, and the diameter of SF networks is smaller than the diameter of ER networks with the same average degree. The fact that $l_\T{max}+1\geq\tau_\T{I}^\T{c}$ suggest that $\tau_\T{I}^\T{c}\sim\log(N)$. {\bf (b)}~At the critical point $c_\T{II}=1.0$ the diameter of ER networks scales as $D\sim N^{1/3}$, suggesting that $\tau_\T{I}^\T{c}$ scales as a powerlaw of $N$. {\bf (c)}~Above the critical point ($c_\T{II}=4.0>1$) there is no direct connection between $D$ and $\tau_\T{I}^\T{c}$, nonetheless observations suggest $\tau_\T{I}^\T{c}\sim \log N$. In contrast with the $c_\T{II}\leq 1$ case, $\tau_\T{I}^\T{c}$ increases more rapidly for SF-SF networks than for ER-ER networks.
	Each data point is the average over $100$ randomly generated networks with $c_\T{I}=4.0$ and error bars represent the standard deviation.}
	\label{fig:secIIIC_N}
\end{figure}

Both $l_\T{max}$ and $n_\T{cycle}$ only depend on Layer II, furthermore both strongly depend on whether Layer II contains a strongly connected component (SCC) or not. Uncorrelated random directed networks -- both ER and SF -- undergo a percolation transition at $c=1$~\cite{SCH02}. If $c< 1$, the network is composed of small tree components, meaning the $n_\T{cycle}=0$ and $l_\T{max}$ is equal to the diameter $D$ of the network. If the system is in the critical point $c=1$, the size of the largest component $S$ diverges as $N\rightarrow \infty$, but the relative size $S/N$ remains zero. The largest component contains a small number of cycles; therefore $D$ is only approximately equal to $l_\T{max}$.  If $c>1$, a unique giant SCC emerges which contains cycles; therefore $n_\T{cycle}>0$ and $l_\T{max}$ is no longer directly connected to the diameter. Rigorous mathematical results show that the diameter of the ER model scales as $D\sim \log(N)$ for $c\neq 1$, and $D\sim N^{1/3}$ for $c=1$, the latter corresponding to percolation transition point~\cite{NAC08}, suggesting that the critical timescale parameter $\tau_\T{I}^\T c$ also depends on $N$. Indeed, Figure~\ref{fig:secIIIC_N} shows that $\tau_\T{I}^\T{c}$ monotonically increases with $N$ for both ER-ER and SF-SF networks. 

We now scan possible values of $c_\T{I}$ while keeping $c_\T{II}$ and $N$ fixed, we find that $n_\T i(c_\T{I})$  and $\tau_\T{I}^\T{c}(c_\T{I})$ quickly converges to its respective lower and upper bound provided by Eqs.~(\ref{eq:tauI_ni_upperbound}) and (\ref{eq:tauIc_upperbound}) (Fig.~\ref{fig:secIIIC}b-c). Varying $c_\T{II}$ and keeping  $c_\T{I}$ fixed shows more intricate behavior: $\tau_\T{I}^\T{c}(c_\T{II})$ increases, peaks and decreases again (Fig.~\ref{fig:secIIIC}d). This is explained by changes in the structure of Layer II: For small $c_\T{II}$ the network is composed of small components with tree structure, increasing $c_\T{II}$ agglomerates these components, thus increasing $l_\T{max}$. For large $c_\T{II}$, a giant SCC exists supporting many cycles, as $c_\T{II}$ increases more and more nodes can be covered with cycles reducing $l_\T{max}$. At the critical point $c_\T{II}^*=1$ the giant SCC emerges, and the largest component consists of $N^\alpha$ nodes ($0<\alpha<1$) with only few cycles, providing the peak of $\tau_\T{I}^\T{c}(c_\T{II})$. Although $c_\T{II}^*=1$ for both ER and SF networks in the $N\rightarrow\infty$ limit, finite size effects delay the peak of $\tau_\T{I}^\T{c}$ for SF-SF networks. Below the transition point, $\tau_\T{I}^\T{c}$ is smaller for ER-ER networks than for SF-SF networks with the same average degree. In contrast, above the transition point SF-SF networks have larger $\tau_\T{I}^\T{c}$. A likely explanation is that the cycle cover of SF networks is smaller than the cycle cover of ER networks with the same average degree, thus more nodes can potentially participate in the longest control path that does not involve cycles.

The number of inputs above the critical timescale parameter $n_\T i(\tau_\T I=\infty)$ is also affected by the cycle cover of Layer II (Fig.~\ref{fig:secIIIC}e): For $c_\T{II}<1$, Layer II does not contain cycles yielding $n_\T i(\tau_\T I=\infty)=1$; for large $c_\T{II}$, Layer II can be completely covered with cycles, and $n_\T i(\tau_\T I=\infty)$ is determined by $n_i^\T I$, the number of inputs needed to control Layer I in isolation.

In summary, if Layer II updates faster, timescale separation reduces controllability up to a critical timescale parameter $\tau^\T c_\T{I}$. For the model networks, the value of $\tau^\T c_\T{I}$ depends on whether Layer II has a giant SCC; $\tau^\T c_\T{I}$ has the highest value at the percolation threshold of Layer II. If Layer II does not contain a giant SCC, degree heterogeneity decreases $\tau^\T c_\T{I}$; above the percolation threshold homogeneous networks have lower $\tau^\T c_\T{I}$. For all timescale parameters, it remains true that ER-ER networks require less independent inputs than SF-SF networks with the same average degree.

\section{Conclusions}\label{sec:conclusions}

Here we explored controllability of interconnected complex systems with a model that incorporates common properties of these systems: (i)~it consists of two layers each described by a complex network; (ii)~the operation of each layer is characterized by a different, but often comparable timescale and (iii)~the external controller only interacts with one layer directly.
{We focused on two-layer multiplex networks, meaning that we assume one-to-one coupling between the nodes of the two layers. Our motivation for this choice was to ensure analytical tractability and to isolate the specific role of timescales from the effect of more complex multilayer network structure. Results obtained for more general multilayer networks will ultimately be shaped by a variety of features such as complex interconnectivity structure, correlations in network structure and details of dynamics. However, even by studying multiplex networks, we uncovered nontrivial phenomena, attesting that without understanding each individual effect, it is impossible to fully understand a system as a whole.}

Using structural controllability we were able to solve the model, thereby directly linking controllability to a graph combinatorial problem. We investigated the effect of network structure and timescales by measuring the minimum number of independent inputs needed for control, $N_\T i$. Overall we found that dense networks with homogeneous degree distribution require less inputs than sparse heterogeneous networks, in line with previous results for single-layer networks~\cite{LIU11}. We showed that if we control the faster layer directly, $N_\T i$ decreases with increasing timescale difference, but only up to a critical value. Above the critical timescale difference, $N_\T i$ is completely determined by the faster layer and we do not have to take into account the multiplex structure of the system. This critical timescale separation is expected to be large if the faster layer would be easy to control and the slower layer would be hard to control in isolation. If we interact with the slower layer, control is increasingly difficult for increasing timescale difference, again up to a critical value, above which $N_\T i$ still depends on the structure of both layers. In this case the critical timescale difference largely depends on the longest control path that does not involve cycles in the faster layer.

Although our model offers only a stylized description of real systems, it is a tractable first step towards understanding the role of timescales in control of interconnected networks. {By identifying the network characteristics that affect important measures of controllability, such as minimum number of inputs needed for control and critical timescale difference, our results serve as a starting point for future work that aims} to relax some of the model's assumptions. Some of these extensions are relatively straightforward using the tool set developed here, for example, the effect of higher order network structures can be studied by adding correlations to the underlying networks. Other extensions are more challenging, e.g. if the interconnection between the layers is incomplete or the layers contain different number of nodes, the minimum input problem is computationally more difficult; therefore investigating such systems would require development of efficient approximation schemes.  {Structural control theory does not take the link weights into account; therefore answering questions that depend on the specific strength of the connections require the development of different tools. For example, for continuous-time systems the timescales are encoded in the strength of the interactions; or the minimum control energy also depends on value of the link weights.}

\section*{Acknowledgements}
We thank Yang-Yu Liu, Philipp H\"ovel and Zs\'ofia P\'enzv\'alt\'o for useful discussions. We gratefully acknowledge support from the US Army Research Office Cooperative Agreement No. W911NF-09-2-0053 and MURI Award No. W911NF-13-1-0340, and the Defense Threat Reduction Agency Basic Research Awards HDTRA1-10-1-0088 and HDTRA1-10-1-00100.

\appendix

\section{Analytical solution for \textnormal{$\tau_\T{I}=\tau_\T{II}=1$}}
\label{app:sec:analytical}

In this section we derive an analytical solution of $n_\T i = N_\T i /N$ in case of $\tau_\T{I}=\tau_\T{II}=1$ for two-layer random networks with predefined degree distribution as defined in Sec.~\ref{sec:results}. This network model is treelike in the $N\rightarrow\infty$ limit; therefore it lends itself to the generating function formalism. The approach described here is based on calculating the fraction of nodes that are matched in all possible maximum matchings~\cite{JIA13}. This solution is substantially simpler than the one described in Ref.~\cite{LIU11}; however, it only applies to bipartite networks (or to bipartite representations of directed networks), and cannot be generalized to unipartite networks. 

We aim to calculate the expected size of the maximum matching of the following undirected bipartite network $\mathcal B$. Layer I $\mathcal{L}_\T{I}$ and Layer II $\mathcal{L}_\T{II}$ are generated independently either using the ER or the SF model; $V_\T{I}$ and $E_\T{I}$ are the node and link sets of  $\mathcal{L}_\T{I}$ and $V_\T{II}$ and $E_\T{II}$ are the node and link sets of  $\mathcal{L}_\T{II}$. Each node in $v^\T{I}_i\in V_\T{I}$ is split into two copies $v^\T{I}_{i,0}\in V^\T{I}_{0}$ and $v^\T{I}_{i,1}\in V^\T{I}_{1}$, we draw a link $(v^\T{I}_{i,0}-v^\T{I}_{j,1})$ if there exists a link $(v^\T{I}_{i}\rightarrow v^\T{I}_{j})$ in $\mathcal{L}_\T{I}$. We treat $\mathcal{L}_\T{II}$ similarly. We then add links $(v^\T{I}_{i,0}-v^\T{II}_{i,1})$ for all $i$. That is all links in $\mathcal B$ connect exactly one node in $V^\T{I}_{0}\cup V^\T{II}_{0}$ to one node in $V^\T{I}_{1}\cup V^\T{II}_{1}$. Nodes in $V^\T{I}_{0}\cup V^\T{I}_{1}$ belong to Layer I, and nodes in $V^\T{II}_{0}\cup V^\T{II}_{1}$ belong to Layer II. The network $\mathcal B$ is the undirected version of the dynamical graph $\mathcal D_1$ without control signals.

In general, multiple possible maximum matchings may exist in a network. We first calculate the fraction of nodes that are matched in all possible maximum matchings. It was shown in Ref.~\cite{JIA13} that in any network $\mathcal G$ a node $v$ is always matched if and only if at least one of its neighbors is not always matched in $\mathcal G\setminus v$, where  $\mathcal G\setminus v$ is the network obtained by removing node $v$ from $\mathcal G$. We translate this rule to a set of self-consistent equations to calculate the expected fraction of always matched nodes in our random network model in the $N\rightarrow\infty$ limit. We provide comments on the issues of applying the rule proven for finite networks to infinite ones at the end of this section.

To proceed we define a few probabilities. We randomly select a link $e$ connecting two nodes $v^\T{I}_{i,0}\in V^\T{I}_{0}$ and $v^\T{I}_{j,1}\in V^\T{I}_{1}$. Let $\theta^\T{I}_{0}$ be the probability that $v^\T{I}_{i,0}$ is always matched in  $\mathcal B\setminus e$, and $\theta^\T{I}_{1}$ be the probability that $v^\T{I}_{j,1}$ is always matched in $\mathcal B\setminus e$. Similarly we randomly select a link $e$ connecting a node $v^\T{I}_{i,0}\in V^\T{I}_{0}$ with a node $v^\T{II}_{i,1}\in V^\T{II}_{1}$. Let $\theta^\T{I,II}_{0}$ be the probability that node $v^\T{I}_{i,0}$ is always matched in $\mathcal B\setminus e$, and $\theta^\T{I,II}_{1}$ be the probability that node $v^\T{II}_{i,1}$ is always matched in  $\mathcal B\setminus e$. The probabilities $\theta^\T{II}_{0}$ and $\theta^\T{II}_{1}$ are defined similarly. According to the rule described above these quantities can be determined by the following set of equations:
\begin{equation}
\label{eq:app:theta}
\begin{split}
\theta^\T{I}_{0} &= 1 - H^\T{I}(\theta^\T{I}_{1})\theta^\T{I,II}_{1}, \\
\theta^\T{I}_{1} &= 1 - H^\T{I}(\theta^\T{I}_{0}),\\
\theta^\T{I,II}_{0} &= 1 - G^\T{I}(\theta^\T{I}_{1}),\\
\theta^\T{I,II}_{1} &= 1 - G^\T{II}(\theta^\T{II}_{0}),\\
\theta^\T{II}_{0} &= 1 - H^\T{II}(\theta^\T{II}_{1}),\\
\theta^\T{II}_{1} &= 1 - H^\T{II}(\theta^\T{II}_{0})\theta^\T{I,II}_{0},
\end{split}
\end{equation}
where $G^\T{I/II}(x)=\sum_{k=0}^\infty P^\T{I/II}(k)x^k$ are the generating functions of the degree distributions and $H^\T{I/II}(x)=\sum_{k=1}^\infty k/\avg{k} P^\T{I/II}(k)x^{k-1}$ are the generating functions of the excess degree distributions.

If we remove a node $v$ which is not always matched, the size of the maximum matching does not decrease. However, if $v$ is matched in all maximum matchings, the number of matched nodes will decrease by two. Therefore to count the size of the maximum matching, we first count the number of nodes that are always matched. By doing so, we have double counted the case when an always matched node is matched by another always matched one. This case occurs for each link $e$ that connects two nodes that are not always matched in $\mathcal G\setminus e$. Combining these two contributions, the expected number of links in the matching is
\begin{equation}
\begin{split}
N_\T{match} =& N[1 - G^\T{I}(\theta^\T{I}_{1})\theta^\T{I,II}_{1}] + N[1 - G^\T{I}(\theta^\T{II}_{0})] + N[1 - G^\T{II}(\theta^\T{I}_{1})] + N[1 - G^\T{II}(\theta^\T{II}_{0})\theta^\T{I,II}_{0}] -\\
 -& c_\T{I}N(1-\theta^\T{I}_{0})(1-\theta^\T{I}_{1}) - N(1-\theta^\T{I,II}_{0})(1-\theta^\T{I,II}_{1}) - c_\T{II}N(1-\theta^\T{II}_{0})(1-\theta^\T{II}_{1}),
\end{split}
\end{equation}
where the first four terms count the number of nodes that are always matched in  $V^\T{I}_{0}$, $V^\T{I}_{1}$,$V^\T{II}_{0}$ and $V^\T{II}_{1}$, respectively; and the last three terms correct the double counting. The expected number of independent inputs needed is determined by the number of unmatched nodes in $V^\T{I}_{1}$ and $V^\T{II}_{1}$:
\begin{equation}
N_\T i = 2N - N_\T{match}.
\end{equation}
Due to the links between Layer I and Layer II, the size of the maximum matching is at least $N$, meaning that $N_\T i\leq N$. Therefore we normalize $N_\T i$ by $N$, yielding
\begin{equation}
\begin{split}
n_\T i = & G^\T{I}(\theta^\T{I}_{1})\theta^\T{I,II}_{1} + G^\T{I}(\theta^\T{II}_{0}) + G^\T{II}(\theta^\T{I}_{1}) + G^\T{II}(\theta^\T{II}_{0})\theta^\T{I,II}_{0} - 2 +\\
 +& c_\T{I}(1-\theta^\T{I}_{0})(1-\theta^\T{I}_{1}) + (1-\theta^\T{I,II}_{0})(1-\theta^\T{I,II}_{1}) + c_\T{II}(1-\theta^\T{II}_{0})(1-\theta^\T{II}_{1}).
\end{split}
\end{equation}

\subsubsection*{Comments on matchings in the configuration model}
\label{sec:app:infintematching}

The method we described to calculate the expected size of the maximum matching does not work for unipartite ER or SF networks generally. The reason for this is that above a critical average degree $c^*$ a densely connected subgraph forms, which is referred to as the core of the network (sometimes leaf removal core or computational core)~\cite{BAU01,COR06,LIU12}. To derive Eq.~(\ref{eq:app:theta}), we assume that the neighbors of a randomly selected node $v$ are independent of each other in $\mathcal B\setminus v$ and removing a single node does not influence macroscopic properties, e.g. $\theta$. The effect of the core is that these assumptions no longer hold and removing just a few nodes may drastically change the number of always matched nodes. Possible way of circumventing this problem is to introduce a new category of nodes: in addition to keeping track of nodes that are sometimes matched and always matched, we separately account for nodes that are almost always matched~\cite{ZDE06}.

The reason why the calculation works for bipartite networks is that a core in the bipartite network will have two sides: all nodes on one side will be always matched and all nodes on other will be some times matched~\cite{JIA13,JIA14,POS14}. If the expected size of the core on the two sides is different, finite removal of nodes will not change macroscopic properties. If the expected size of the two sides of the core is the same, removal of finite nodes may change which side is always matched and which side is sometimes matched~\cite{JIA13}. However, this does not change expected fraction of matched nodes; therefore does not interfere with the calculations.

\section{Algorithms}
\label{app:sec:algorithms}

\subsection{Cycle cover (\textnormal{$N_\T{cycle}$})}
\label{app:sec:ncycle}

To find the maximum cycle cover of a directed network $\mathcal L$, we assign weight $0$ to each link in $\mathcal L$; and we add a self-loop with weight $1$ to each node that does not already have a self-loop. Then we find the minimum weight maximum directed matching in $\mathcal L$ augmented with self-loops by converting the problem to a minimum cost maximum flow problem. The maximum matching is guaranteed to be perfect, because each node has a self-loop. The minimum weight perfect matching in the directed network corresponds to a perfect cycle cover where the number of self-loops with weight $1$ is minimized. Therefore the maximum cycle cover in $\mathcal L$ without extra self-loops is
\begin{equation}
N_\T{cycle}= N-W,
\end{equation}
where $W$ is the sum of the weights of the links in the minimum weight perfect matching.

\subsection{Longest control path not involving cycles (\textnormal{$l_\T{max}$})}
\label{supp:sec:lmax}

In this section we provide the algorithm to measure the longest control path not involving cycles $l_\T{max}$ of Layer II of a two-layer network for the case $\tau_\T{I}\geq 1$ and $\tau_\T{II}=1$. The algorithm itself serves as the precise definition of $l_\T{max}$.

Given a two-layer directed network $\mathcal M$, let $N_\T{cycle}$ be the maximum number of nodes that can be covered by node disjoint cycles in Layer II. To measure $l_\T{max}$, first we construct the dynamical graph $\mathcal D_{l}^\T{II}$ representing the time evolution of the Layer II between time $t=0$ and $t=l$ as if it would be isolated as defined in Sec.~\ref{sec:single-layer_net}. We search for disjoint control paths connecting nodes at time step $t=0$ with nodes at time step $t=l$, e.g. each control path connects a node $v^\T{II}_{i,0}$ with $v^\T{II}_{j,l}$. The maximum number of such paths $N_\T{path}(l)$ provides the maximum number of internally controlled nodes if $\tau_\T{I}=l$. To determine $N_\T{path}(l)$ we convert the problem to a maximum flow problem: We set the capacity of each link and each node in $\mathcal D_{l}^\T{II}$ to 1. We then find the maximum flow connecting source node set $V^\T{II}_0=\{v^\T{II}_{i,0}\vert i=1,2,\ldots,N\}$ to target node set $V^\T{II}_l=\{v^\T{II}_{i,l}\vert i=1,2,\ldots,N\}$ using a maximum flow algorithm of choice. The maximum flow provides $N_\T{path}(l)$. And $l_\T{max}$ is defined as one less than the smallest value of $l$ such that
\begin{equation}
N_\T{path}(l)=N_\T{cycle}.
\end{equation}

\begin{figure}[H]
	\centering
	\includegraphics[scale=.55]{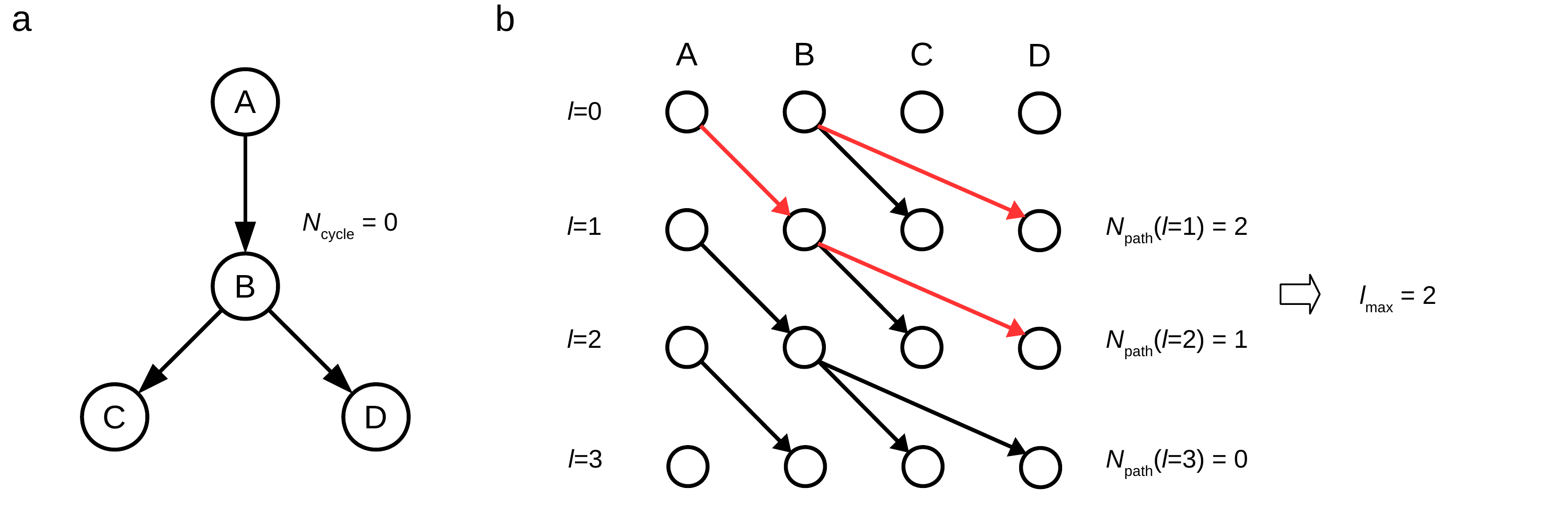}
	\caption{$l_\T{max}${ \bf -- Example 1.} {\bf (a)}~A directed network with tree structure; therefore not containing cycles. The diameter $D=2$ is the length of the longest path. {\bf (b)}~We count the maximum number of disjoint control paths $N_\T{path}(l)$ which connect nodes at time step $0$ with nodes at time step $l$. We find that $l^\prime=3$ is the smallest value of $l$ such that $N_\T{path}(l)=N_\T{cycle}=0$; therefore $l_\T{max}=2$. There are no cycles; therefore $l_\T{max}=D$.}
	\label{app:fig:lmax-example-1}
\end{figure}

\begin{figure}[H]
	\centering
	\includegraphics[scale=.55]{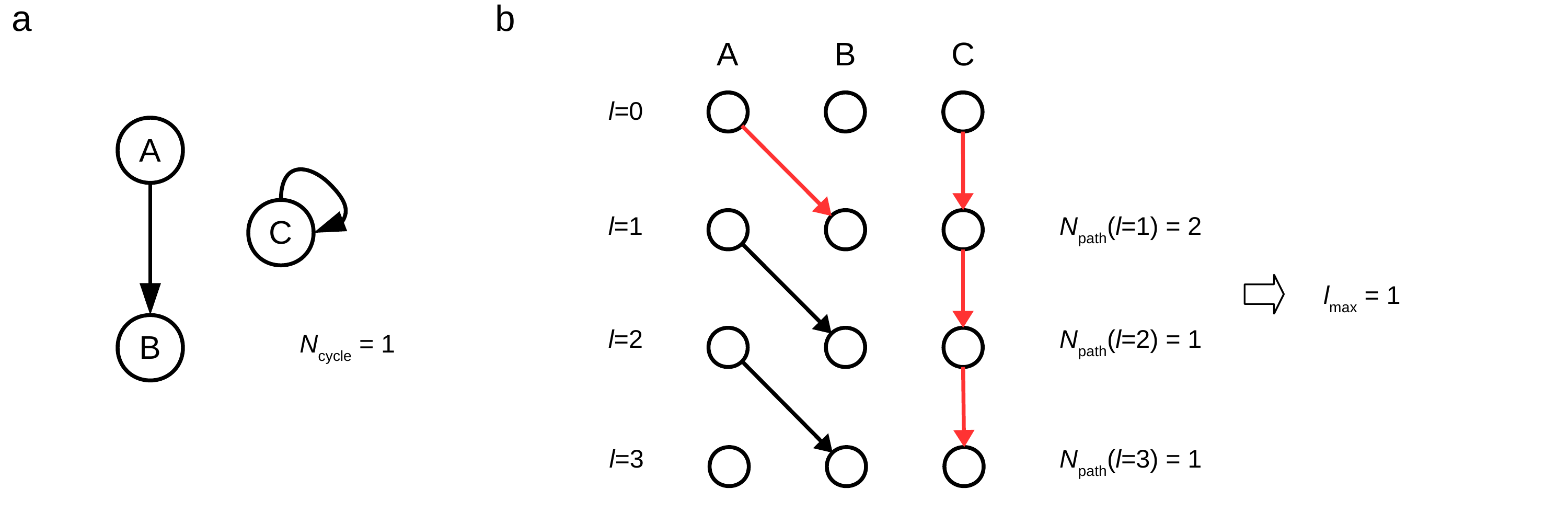}
	\caption{$l_\T{max}${ \bf -- Example 2.} {\bf (a)}~A directed network containing a cycle. The size of the maximum cycle cover is $N_\T{cycle}=1$. {\bf (b)}~We count the maximum number of disjoint control paths $N_\T{path}(l)$ which connect nodes at time step $0$ with nodes at time step $l$. We find that $l^\prime=2$ is the smallest value of $l$ such that $N_\T{path}(l)=N_\T{cycle}=1$; therefore $l_\T{max}=1$. $N_\T{path}(l)$ remains non-zero for $l>l_\T{max}$, showing that cycles can support control paths of any length.}
	\label{app:fig:lmax-example-2}
\end{figure}

Figures~\ref{app:fig:lmax-example-1} and \ref{app:fig:lmax-example-2} provide two examples to illustrate the calculation of $l_\T{max}$.


%

\newpage

\end{document}